\documentclass[runningheads]{llncs}

\usepackage{eccv}
\usepackage{eccvabbrv}
\usepackage{graphicx}
\usepackage{booktabs}
\usepackage{multirow}
\usepackage[accsupp]{axessibility}
\usepackage{hyperref}
\usepackage{orcidlink}

\graphicspath{{figures/}}
\begin{document}

\title{Efficient Passive Acoustic Monitoring of Killer Whales Using a Two-Stage
Detection and Ecotype Classification Cascade}
\titlerunning{Efficient Passive Acoustic Monitoring of Killer Whales}

\author{Daniela Ruiz\inst{1,2}\orcidlink{0000-0001-6636-1173} \and
Manuel Castellote\inst{1}\orcidlink{0009-0002-9306-5577} \and
Zhongqi Miao\inst{1}\orcidlink{0000-0002-0439-8592} \and
Carl Chalmers\inst{1}\orcidlink{0000-0003-0822-1150} \and
Bruno Demuro\inst{1}\orcidlink{0000-0001-9922-8811} \and
Rahul Dodhia\inst{1}\orcidlink{0000-0003-3812-3906} \and Pablo Arbelaez\inst{2}\orcidlink{0000-0001-5244-2407} \and Juan M. Lavista\inst{1}\orcidlink{0000-0002-9654-3178}}
\authorrunning{D.~Ruiz et al.}
\institute{Microsoft AI for Good Research Lab, Redmond, Washington, USA \\ 
\email{\{v-druizlopez,v-manoloc,zhongqimiao,chalmerscarl,\\ 
v-brunode,radodhia,jlavista\}@microsoft.com}  \and
Universidad de los Andes, Bogota, Colombia \\ 
\email{\{da.ruizl1,pa.arbelaez\}@uniandes.edu.co}}

\maketitle

\begin{abstract}
Passive acoustic monitoring of killer whales is particularly important for conservation of the endangered Southern Resident killer whale population, but requires accurate models that can operate in real time under severe class imbalance and deployment shift. We propose a lightweight ResNet-based two-stage cascade that first detects killer whale vocalizations and then classifies confident detections into five eastern North Pacific ecotypes, abstaining on ambiguous calls. We train and evaluate the pipeline on the DCLDE 2027 dataset, where the detector achieves 0.960 macro-F1 and the classifier 0.958, outperforming frozen Perch 2.0 embeddings on the five-ecotype benchmark. By separating detection from ecotype recognition, the end-to-end cascade improves seven-class macro-F1 from 0.919 for a single-stage model to 0.933, with the largest gain on the rare OKW ecotype. To assess transfer beyond the benchmark, we use active learning to adapt the Stage~1 to the acoustic environment of Puget Sound, WA, increasing killer whale detection F1 from 0.405 to 0.755 on manually verified detection windows. Finally, each stage processes a 3~s window in approximately 1.4 ms on an NVIDIA H100, enabling faster than real time inference. These results demonstrate that the proposed two-stage cascade pipeline enables reliable killer whale detection and classification, adaptation to new acoustic domains, and real-time monitoring for conservation applications.


\end{abstract}

\section{Introduction}
\label{sec:introduction}

Killer whales (\textit{Orcinus orca}) are globally distributed apex predators. In the eastern North Pacific Ocean, their populations are divided into acoustically distinct ecotypes: Southern Residents (SRKW), Northern Residents (NRKW), Southern Alaska Residents (SAR), mammal-eating Transients/Bigg's (TKW) and Offshore (OKW). SRKW population is listed as Endangered under the U.S.\ Endangered Species Act (ESA)~\cite{noaa2026}, making reliable monitoring a conservation priority. The results of SRKW occurrence in near real-time are critical for informed seasonal management actions, such as the speed and approach regulations of the vessel and depend on timely whale alerts~\cite{scott2024}. Improving the speed and reliability of these monitoring systems can therefore have an immediate conservation impact.

Passive acoustic monitoring (PAM) is the only monitoring modality capable of providing continuous, weather-independent, non-invasive observations across the large spatial and temporal scales occupied by these populations. However, modern hydrophone networks produce thousands of hours of recordings every year, making manual annotation infeasible. Automatic detection and ecotype classification are therefore essential components of scalable monitoring pipelines.

Although several public datasets and benchmarks have recently accelerated research in this direction, models evaluated on curated benchmarks face additional challenges when deployed on continuous acoustic streams. Real-world PAM recordings are overwhelmingly dominated by background noise with only a small fragment of these containing signals of interest. Consequently, annotated datasets substantially underestimate the prevalence and diversity of background sounds encountered during deployment. A model operating on edge devices or a cabled hydrophone system must therefore maintain high recall to avoid missing killer whale vocalizations while achieving sufficiently high precision to minimize the human effort required to verify predictions. Furthermore, deployment on autonomous hydrophone platforms imposes strict computational, power, and memory constraints, favoring lightweight models that can operate in real time.

To address these deployment challenges, we propose a lightweight two-stage acoustic recognition pipeline suitable for real-time passive acoustic monitoring. A first stage detects killer whale vocalizations from continuous hydrophone recordings, while a second stage classifies confidently detected vocalizations into the five eastern North Pacific ecotypes, with an abstention mechanism for acoustically ambiguous vocalizations. Both stages operate on mel spectrograms using compact ResNet-18~\cite{he2016} backbones, providing a unified and lightweight architecture. To demonstrate adaptability under domain shift, we fine-tune the detector using site-specific data from an ongoing edge deployment in Puget Sound, WA, tailoring it to the target acoustic environment.

\section{Related Work}

Treating mel spectrograms as images has become the dominant paradigm for supervised bioacoustics detection and classification. BirdNET~\cite{kahl2021} established this paradigm at continental scale, demonstrating that image-classification architectures applied to mel spectrograms enable accurate large-scale bird monitoring. While the original BirdNET architecture was based on ResNet~\cite{he2016}, subsequent releases of BirdNET-Analyzer have adopted EfficientNet backbones, reflecting the continued effectiveness of lightweight CNNs for large-scale passive acoustic monitoring. Perch~\cite{ghani2023} extended this paradigm beyond task-specific classification by learning transferable bioacoustic representations from a large, taxonomically diverse training corpus, enabling downstream adaptation across multiple acoustic monitoring tasks. More specialized supervised models include SurfPerch~\cite{williams2025using}, which targets marine bioacoustics, and MetaPerch~\cite{chasmai2026metaperch}, which incorporates temporal and geographic metadata into bioacoustic classification.

More recently, large-scale pretrained audio encoders have emerged as an alternative to supervised models. General-purpose foundation models such as BEATs~\cite{chen2022beats}, AudioMAE~\cite{huang2022}, and EAT~\cite{chen2024eat} learn transferable audio representations from large unlabeled corpora using self-supervised learning. Bioacoustic-specific models extend the same paradigm to animal communication across multiple taxa; examples include AVES~\cite{hagiwara2022aves}, Bird-MAE~\cite{rauch2025mae}, animal2vec~\cite{zimmermann2026}, and AVEX~\cite{miron2026}. Rather than directly predicting species labels, these models produce embeddings that can be adapted to downstream tasks through linear probes or lightweight fine-tuning.

Although self-supervised representation learning has become the dominant paradigm in computer vision and increasingly influences bioacoustics, Perch 2.0~\cite{vanmerrienboer2026} demonstrates that large-scale supervised learning remains highly competitive. Trained on millions of labeled recordings spanning 14,597 taxa, Perch 2.0 substantially expands the original Perch model and achieves state-of-the-art performance on BirdSet~\cite{rauch2025birdset}, a standardized benchmark for evaluating bird sound classification across diverse datasets, and BEANS~\cite{hagiwara2022beans}, a benchmark designed to assess the generalization of bioacoustic foundation models across animal taxa and downstream tasks.

Early work on automatic killer whale recognition has focused primarily on binary detection. ORCA-SPOT~\cite{bergler2019} proposed a modified ResNet-18~\cite{he2016} architecture for detecting killer whale vocalizations from spectrograms, showing that removing the initial max-pooling layer preserves fine temporal-frequency structure important for high-frequency pulsed calls. More recent work has addressed ecotype classification in the Pacific Northwest using the DCLDE dataset~\cite{palmer2025}. Transfer learning from BirdNET demonstrated accurate discrimination between SRKW and TKW, although performance decreased on data collected using a different hydrophone system \cite{palmer2026}. On the broader five-ecotype benchmark, frozen Perch 2.0 embeddings combined with a linear classifier were evaluated on the same task, demonstrating that large-scale bioacoustic foundation models transfer effectively to underwater acoustic monitoring~\cite{burns2025}. 

\section{Dataset}
\label{sec:dataset}

\subsection{Source Corpus}

We train and evaluate our approach on the largest publicly available dataset for killer whale passive acoustic monitoring, released for the 2027 Biennial Conference and Workshop on Detection, Classification, Localization, and Density Estimation of Marine Mammals using Passive Acoustics (DCLDE)~\cite{palmer2025}.

The original corpus comprises 13,084 audio recordings collected from 32 hydrophone deployments operated by 10 data providers and 27 recording systems across the eastern North Pacific Ocean. Recordings span a wide range of acoustic environments, deployment depths (8--1,400~m), sampling rates (16--256~kHz), and annotation protocols. The dataset totals 1,065.8 hours of audio, of which only 42.5 hours (4.0\%) are annotated. Annotations were provided at call, detection, or file levels, with PAMGuard-assisted preprocessing used for several subsets to identify candidate detections for verification~\cite{palmer2025}. Thus, unannotated portions of recordings may contain detections that were missed or not labeled. The labeled data consists of 207,574 annotations across four coarse acoustic categories: Killer Whale (KW), Humpback Whale (HW), Undetermined Biological sounds (UndBio), and Abiotic sounds (AB). Of the 58,261 KW annotations, 52,196 (89.6\%) carry an ecotype label and are distributed across: SRKW (40\%), TKW (24\%), NRKW (16\%), SAR (15\%), and OKW (5\%).

Before training, we applied three preprocessing steps to improve annotation consistency. First, we removed file-level annotations that did not provide temporal localization of signals within files. Second, we discarded KW annotations marked as uncertain when confidence labels were available. Third, we removed zero-duration annotations. These filtering steps removed only 1,532 annotations (0.74\%), yielding a final dataset of 206,042 annotated events. DCLDE annotations distribution is shown in \Cref{tab:distribution}.

\begin{table}[t]
\centering
\caption{DCLDE annotations and window-level training dataset.}
\label{tab:distribution}
\small
\begin{tabular}{lrrrr}
\toprule
Class & DCLDE annotations & \% & Windows & \% \\
\midrule
SRKW       & 20,447  & 10\% & 82,935  & 14\% \\
TKW        & 12,707  & 6\%  & 51,402  & 8\% \\
NRKW       & 8,266   & 4\%  & 37,772  & 6\% \\
SAR        & 8,078   & 4\%  & 33,823  & 6\% \\
OKW        & 2,698   & 1\%  & 11,008  & 2\% \\
Unassigned & 6,065   & 3\%  & 23,534  & 4\% \\
\midrule
\textbf{KW} & \textbf{58,261} & \textbf{28\%} & \textbf{240,474} & \textbf{39\%} \\
\midrule
HW         & 124,977 & 61\% & 126,922 & 21\% \\
UndBio     & 11,819  & 6\%  & 45,365  & 7\% \\
\midrule
\textbf{Bio} & \textbf{136,796} & \textbf{66\%} & \textbf{172,287} & \textbf{28\%} \\
\midrule
AB          & 10,985 & 5\% & 36,220  & 6\% \\
Internal BN & 0      & 0\% & 30,720  & 5\% \\
External BN & 0      & 0\% & 133,060 & 22\% \\
\midrule
\textbf{NonBio} & \textbf{10,985} & \textbf{5\%} & \textbf{200,000} & \textbf{33\%} \\
\midrule
\textbf{Total} & \textbf{206,042} & \textbf{100\%} & \textbf{612,761} & \textbf{100\%} \\
\bottomrule
\end{tabular}
\end{table}

\subsection{Windows and Spectrogram Representation}
\label{sec:windows}

To obtain a consistent input representation across the heterogeneous DCLDE corpus, all recordings are first resampled to a common sampling rate of 24~kHz. Although the original recordings span sampling rates from 16 to 256~kHz, analysis of the annotated frequency bounds shows that 98.4\% of vocalizations have an upper frequency at or below the corresponding 12~kHz Nyquist limit. This resolution therefore preserves nearly all information relevant to KW detection, including the lower frequencies associated with echolocation, as well as the full frequency range of social signals relevant to ecotype classification, while avoiding the computational cost of processing higher sampling rates.

Training windows are generated directly from the filtered annotations. Rather than exhaustively segmenting entire sound files, training samples are generated only around annotated events. Each sample consists of a fixed 3~s audio window, which was selected based on the annotation duration distribution: over 99\% of labeled events are shorter than 3~s, ensuring that a complete vocalizations can be contained within a single window while preserving sufficient acoustic context. Because HW annotations already dominate the dataset (\cref{tab:distribution}), a single 3~s window centered on each HW annotation is extracted. In contrast, KW, AB, and UndBio annotations are expanded into five overlapping windows to compensate for class imbalance in the downstream detection task. The central window is centered on the annotation midpoint, while two additional windows are shifted by \(\pm1.5\)~s and two more by \(\pm3.0\)~s, providing approximately 9~s of temporal context around each annotated event.

Each generated window is assigned an annotation status based on its temporal overlap with reference annotations. Given that the shortest KW annotation in dataset is 0.1~s, windows with at least 0.1~s overlap are considered annotated, those with no overlap unannotated, and those with shorter overlaps or extending beyond recording boundaries are discarded to avoid ambiguous supervision. Since most annotations are shorter than the 3~s analysis window, the three central windows typically overlap the annotated event and inherit its label, whereas the outer windows remain unannotated. This procedure augmented KW, AB, and UndBio samples and generated 69,835 unannotated windows.

To enrich the training dataset with background noise (BN), we initially considered using the unannotated windows. However, preliminary inspection showed that some contained KW or HW vocalizations. We therefore screened the 69,835 candidates with the pretrained binary Whale/No Whale detector from~\cite{castellote2026}, retaining 30,720 background noise windows with maximum sigmoid confidence below 0.30. As the BN samples remained limited relative to the other classes, we additionally selected 133,060 3 s background noise intervals from the Cook Inlet beluga monitoring corpus used in~\cite{castellote2026}. Although these data were collected in a subarctic environment of the eastern North Pacific, their background noise characteristics are broadly representative of estuarine and coastal habitats. Accordingly, we map the original DCLDE classes to three detection classes: KW, Bio (HW and UndBio), and NonBio (AB, internal BN, and external BN). The resulting training distribution is summarized in the Windows column of \cref{tab:distribution}. 

Finally, each audio window is converted to a log-mel spectrogram using a 1024-point FFT, a hop length of 128 samples, and 256 mel frequency bins spanning 200~Hz--12~kHz. Frequencies below 200~Hz are discarded to reduce low-frequency vessel noise while retaining most spectral content of HW whale and complete content for KW vocalizations. Spectrogram magnitudes are converted to the logarithmic decibel scale and clipped to a dynamic range of 80~dB to reduce extreme variability across hydrophone deployments. The resulting fixed-size spectrograms are cached as tensors and used as the common input representation for all models in the proposed pipeline.

\subsection{Dataset Splits}
\label{sec:splits}

\begin{table}[t]
\centering
\caption{Per-class window counts for the data splits in both stages.}
\label{tab:splits}
\small
\begin{tabular}{llrrr}
\toprule
Dataset & Labels & Train & Val & Test \\
\midrule
\multirow{3}{*}{Stage 1 (Detection)}
 & (0) NonBio & 141,080 & 29,721 & 29,199 \\
 & (1) Bio    & 122,736 & 26,288 & 23,263 \\
 & (2) KW   & 168,327 & 36,918 & 35,229 \\
\cmidrule{2-5}
 & \textbf{Total} & \textbf{432,143} & \textbf{92,927} & \textbf{87,691} \\
\midrule
\multirow{5}{*}{Stage 2 (Ecotype)}
 & (0) SRKW & 59,664 & 12,811 & 10,460 \\
 & (1) TKW  & 34,939 & 7,807  & 8,656 \\
 & (2) SAR  & 26,268 & 5,364  & 6,140 \\
 & (3) NRKW & 23,753 & 4,520  & 5,550 \\
 & (4) OKW  & 7,649  & 2,132  & 1,227 \\
\cmidrule{2-5}
 & \textbf{Total} & \textbf{152,273} & \textbf{32,634} & \textbf{32,033} \\
\bottomrule
\end{tabular}
\end{table}

The two stages operate on different label spaces. Stage~1 is trained on the complete window-level dataset comprising 612,761 samples labeled as KW, Bio, or NonBio, whereas Stage~2 is trained only on the 216,940 KW windows carrying an ecotype label. To prevent data leakage, all dataset partitions are generated by grouping windows according to their source sound file, ensuring that windows extracted from the same sound recording never appear in different splits.

As summarized in \cref{tab:splits}, Stage~1 is partitioned into 70\% training, 15\% validation, and 15\% testing while preserving the class distribution of the three detection classes. The Stage~2 test set is then constructed by inheriting every ecotype-labeled recording contained in the Stage~1 test partition, while the remaining recordings are divided into training and validation sets with stratified ecotype distributions. For the end-to-end evaluation described in \cref{subsec:cascade}, we define the cascade test set as the union of all 52,462 NonBio and Bio windows in the Stage~1 test partition and the 32,033 KW windows included in the Stage~2 test set. Consequently, both stages are evaluated on the same held-out recordings, allowing cascade performance to be interpreted in the context of the individual stage performances.

\section{Method}
\label{sec:method}

\subsection{Two-Stage Cascade Pipeline}
\label{subsec:cascade}

\begin{figure}[t]
\centering
\includegraphics[width=\textwidth]{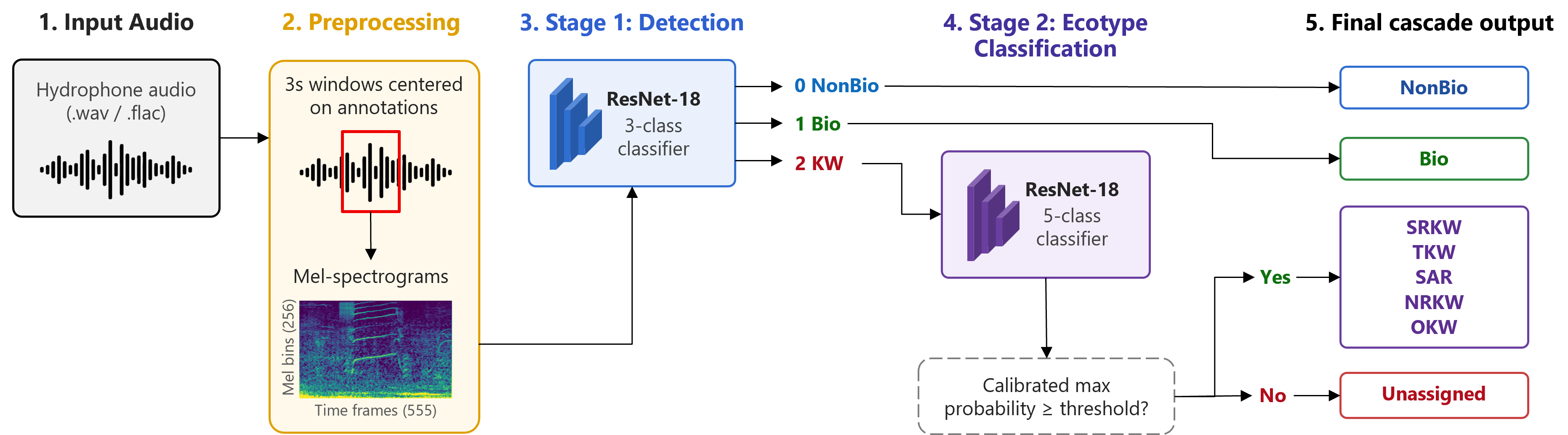}
\caption{Pipeline overview. Sound recordings are preprocessed into 3-second windows centered on annotations and converted to mel spectrograms. Stage~1 detects the presence of KWs, separating them from biological and non-biological sounds. Only KW detections are forwarded to the ecotype classifier. Predictions whose calibrated confidence falls below the threshold are reported as Unassigned; otherwise, the corresponding ecotype is returned.}
\label{fig:pipeline}
\end{figure}

\Cref{fig:pipeline} illustrates the proposed two-stage acoustic recognition pipeline. Each 3-s sound window is first converted into a log-mel spectrogram following the preprocessing procedure described in \cref{sec:windows}. The first stage detects KW vocalizations by classifying each spectrogram into one of three categories: NonBio, Bio, or KW. Predictions classified as NonBio or Bio terminate the pipeline, while only windows predicted as KW are forwarded to the second stage.

The second stage performs ecotype classification, assigning detected KW vocalizations to one of the five ecotypes: SRKW, TKW, SAR, NRKW, or OKW. Because not every KW vocalization contains sufficient information for reliable ecotype discrimination, Stage~2 is equipped with a confidence-based abstention mechanism. If the calibrated confidence of the predicted ecotype falls below a threshold, the cascade outputs Unassigned instead of forcing an ecotype prediction. Consequently, the complete pipeline predicts one of eight classes: NonBio, Bio, SRKW, TKW, SAR, NRKW, OKW, or Unassigned.

Both stages employ the same lightweight ResNet-18 backbone~\cite{he2016}, enabling a unified architecture throughout the cascade while maintaining computational efficiency for edge deployment. Since the input consists of single-channel mel spectrograms rather than RGB images, the first convolutional layer is modified to accept one input channel by averaging the pretrained ImageNet filters across the RGB dimension. Each network receives a log-mel spectrogram and produces either three detection logits (Stage~1) or five ecotype logits (Stage~2).

\subsection{Training and Computational Efficiency}

Both stages are trained independently using the dataset partitions described in \cref{sec:splits}. The Stage~1 class distribution is nearly balanced, with a 1.4:1 ratio between the largest and smallest classes, so the detector uses standard cross-entropy loss without class weighting. In contrast, the ecotype classification task is highly imbalanced, with SRKW outnumbering OKW by approximately 7.5:1; Stage~2 therefore uses inverse-frequency weighted cross-entropy. Both stages share the same optimization settings: batch size 128, AdamW with a learning rate of \(4\times10^{-4}\) and weight decay of \(10^{-4}\), mixed-precision training, gradient clipping at 1.0, early stopping with a patience of 20 epochs based on validation F1, and a maximum of 50 epochs. Both models were trained on a single NVIDIA H100 NVL GPU.

We benchmark model inference efficiency using cached spectrograms of shape $[B,1,256,555]$, measuring GPU forward-pass latency, throughput, peak allocated memory, and multiply-accumulate operations (MACs). Latency is measured after warm-up over repeated forward passes with CUDA synchronization. Both stages contain approximately 11.2M parameters and require 5.009 GMACs per 3~s window. At batch size 1, Stage~1 and Stage~2 achieve median latencies of 1.419 and 1.412~ms, respectively, corresponding to approximately 705 and 708 windows/s, with 0.064~GiB of peak allocated GPU memory per stage. At batch size 128, throughput reaches approximately 5,120 windows/s per stage, with 1.309~GiB of peak allocated memory. For a 3~s analysis window, the model latency is more than 2,000$\times$ shorter than the window duration, demonstrating faster-than-real-time model execution on this server GPU.

\subsection{Confidence Calibration and Abstention}
\label{sec:calibration}

The ecotype classifier is trained to predict one of five ecotypes, but operational passive acoustic monitoring often includes distant or weak KW vocalizations with insufficient information for reliable ecotype identification. To allow the system to explicitly represent such uncertainty, we calibrate Stage~2 using temperature scaling, estimating the temperature parameter on the validation set and applying it to inference logits before the softmax operation. We then estimate an abstention threshold using the DCLDE datasets from DFO-CRP (West Vancouver Island and Northern Mainland British Columbia sites), whose annotations distinguish KW vocalizations with and without ecotype assignments. We treat ecotype-labeled windows as positive examples and KW windows without ecotype labels as proxies for ambiguous vocalizations. We select the threshold by maximizing by maximizing Youden's \(J\) statistic, \(J=\mathrm{TPR}-\mathrm{FPR}\), on the calibrated confidence scores. We use this criterion rather than binary F1 because the calibration set is dominated by ecotype-labeled vocalizations, which would favor accepting nearly all predictions. During inference, predictions whose maximum calibrated probability falls below this threshold are reported as Unassigned, while the remaining predictions are assigned to the corresponding ecotype.

\subsection{Comparison with Perch}

To contextualize the performance of the proposed ecotype classifier, we compare it with Perch 1.0 and Perch 2.0 using frozen pretrained embeddings. For all three models, embeddings are classified using multinomial logistic regression, with the backbone parameters kept frozen.

We evaluate performance at both the window and recording levels. For window-level evaluation, each input window is independently classified using its corresponding embedding. For recording-level evaluation, window embeddings from each source recording are averaged and L2-normalized before classification with multinomial logistic regression. Recording-level ground-truth labels are assigned by majority vote over the corresponding window labels. For visualization, recording-level t-SNE is computed from the same pooled embeddings and labels.

\subsection{End-to-End Cascade Evaluation}

We evaluate the end-to-end cascade as a seven-class classifier comprising NonBio, Bio, and the five KW ecotypes. The abstention mechanism is disabled because the benchmark contains no ground-truth Unassigned label. Each input window is first processed by the Stage~1 detector, and windows predicted as KW are forwarded to the Stage~2 ecotype classifier.

To characterize errors introduced by the cascade, we separately quantify Stage~1 false negatives, which prevent KW windows from reaching Stage~2, and false positives, which forward non-KW windows to the ecotype classifier. We also assess the potential effect of confidence-based abstention by applying the calibrated threshold described in \cref{sec:calibration} to windows incorrectly forwarded to Stage~2. Finally, we compare the cascade against a single-stage ResNet-18 trained directly on the same seven-class problem using the identical training protocol and evaluation split. 

\subsection{Stage~1 Deployment-Domain Adaptation}

While the DCLDE benchmark provides a standardized evaluation framework, practical conservation systems operate under substantially different acoustic conditions. We therefore evaluate the adaptability of the proposed approach using data from real monitoring conditions in the Puget Sound, WA, USA. Domain adaptation is restricted to the Stage~1 detector, as the ongoing deployment is primarily intended to detect and alert on SRKW presence, and no SRKW detections have been identified. Consequently, insufficient target-domain data are currently available to fine-tune or meaningfully evaluate the Stage~2 ecotype classifier.

The Stage~1 detector experiences a substantial domain shift on Puget Sound data, leading to a marked increase in false-positive detections. Since the vast majority of detections correspond to background noise sources such as tonal components of vessel propulsion noise, maintaining high precision is essential to streamline the near real-time whale alert process and minimize the manual validation effort required from marine mammal experts.

To adapt the detector to the target domain, we fine-tune only the final residual block and the classification head while freezing the remainder of the pretrained ResNet-18 backbone. We augment fine-tuning data by mixing KW vocalizations with vessel recordings at different hydrophone distances to simulate a dominant source of masking and false alarms in Puget Sound. We then use an active learning workflow in which expert-validated KW detections from each deployment round are combined with random samples from the original training data to iteratively update and redeploy the detector. 

We evaluate performance on a held-out set of 1,173 manually verified detection windows collected by the first deployed system over two days of continuous operation. Because audio outside these detections was not reviewed, recall measures retention among verified candidate windows rather than continuous-stream recall. The test set comprises 93.17\% NonBio, 5.71\% KW, and 1.12\% Bio windows. For all models, windows with predicted KW probability above 0.7 are reported as detections. Fine-tuning is focused on reducing deployment-specific false positives while maintaining sensitivity to KW vocalizations.

\section{Results}

\subsection{Stage 1: Detection}

\Cref{tab:stage1} reports the performance of the proposed Stage~1 detector on the held-out test set defined in \cref{sec:splits}. The model achieves 0.960 macro-F1 and 0.992 macro average precision. Performance is balanced across all three classes, with per-class F1 scores ranging from 0.956 (KW) to 0.962 (NonBio). Most detection errors correspond to weak KW vocalizations being classified as background or biologically similar sounds, primarily HW calls, being confused with KW.

We reimplemented the modified ResNet-18 architecture proposed by ORCA-SPOT, which removes the initial max-pooling layer to retain finer time--frequency detail. Under the same training protocol, this modification achieved a macro-F1 of 0.956, compared with 0.960 for the standard ResNet-18, but trained longer. Unlike the original ORCA-SPOT setting, our spectrogram representation contains 555 temporal frames, providing substantially higher temporal resolution than their 128-frame inputs. Consequently, preserving the full spatial resolution after the first convolution provides no measurable benefit in our setting; therefore, we retain the standard ResNet-18 architecture throughout this work.

\begin{table}[t]
\centering
\caption{Performance of the two stages on the DCLDE test set.}
\label{tab:stage-performance}

\begin{subtable}[t]{0.47\textwidth}
\centering
\caption{Stage 1: detection.}
\label{tab:stage1}
\small
\begin{tabular}{@{}lccc@{}}
\toprule
Class & Precision & Recall & F1 \\
\midrule
NonBio & 0.952 & 0.973 & 0.962 \\
Bio    & 0.963 & 0.958 & 0.960 \\
KW   & 0.963 & 0.949 & 0.956 \\
\midrule
Macro  & 0.959 & 0.960 & 0.960 \\
\bottomrule
\end{tabular}
\end{subtable}
\hfill
\begin{subtable}[t]{0.47\textwidth}
\centering
\caption{Stage 2: ecotype classification.}
\label{tab:stage2}
\small
\begin{tabular}{@{}lccc@{}}
\toprule
Ecotype & Precision & Recall & F1 \\
\midrule
SRKW  & 0.953 & 0.982 & 0.967 \\
TKW   & 0.979 & 0.928 & 0.953 \\
SAR   & 0.984 & 0.963 & 0.973 \\
NRKW  & 0.933 & 0.987 & 0.959 \\
OKW   & 0.955 & 0.923 & 0.939 \\
\midrule
Macro & 0.961 & 0.956 & 0.958 \\
\bottomrule
\end{tabular}
\end{subtable}
\end{table}

\begin{table}[t]
\centering
\caption{Window-level comparison with Perch. Values are per-class F1 scores.}
\label{tab:perch}
\small
\begin{tabular}{lrrrrrr}
\toprule
Model & Macro F1 & SRKW & TKW & SAR & NRKW & OKW \\
\midrule
ResNet-18 (ours) & 0.958 & 0.967 & 0.953 & 0.973 & 0.959 & 0.939 \\
Perch v2 + LogReg & 0.933 & 0.965 & 0.948 & 0.963 & 0.938 & 0.851 \\
Perch v1 + LogReg & 0.913 & 0.949 & 0.921 & 0.961 & 0.923 & 0.808 \\
\bottomrule
\end{tabular}
\end{table}

\subsection{Stage 2: Ecotype Classification}

\Cref{tab:stage2} summarizes the performance of the proposed ecotype classifier on the held-out test set comprising 32,033 KW windows. The model achieves a macro-F1 score of 0.958, with balanced performance across all five ecotypes. Every ecotype exceeds an F1 score of 0.93 despite the strong class imbalance present in the training data, with SRKW an SAR achieving the highest performance. OKW, the rarest ecotype, remain the most challenging class but still achieve an F1 score of 0.939.

Temperature scaling with $T=4.54$ reduces the Expected Calibration Error (ECE) from 0.059 to 0.011 on the validation set. The exploratory threshold selected by Youden's $J$ on DFO-CRP data is 0.9344. Perturbing this threshold by $\pm 0.01$ leaves retained-set macro-F1 nearly unchanged (0.996--0.997), but reduces labeled-window coverage from 78.5\% to 74.3\% and ecotype-unassigned proxy acceptance from 67.7\% to 62.2\%. At 0.9344, the model retains 76.6\% of ecotype-labeled windows and increases their retained-set macro-F1 from 0.958 to 0.997. However, 34.5\% of DFO-CRP and 65.0\% of all-source ecotype-unassigned proxies also exceed the threshold. These results show that a single global max-softmax threshold does not reliably separate ecotype-labeled from ecotype-unassigned annotations and should be recalibrated using labeled target-domain data rather than transferred unchanged to a new deployment.

\begin{figure}[t]
\centering
\includegraphics[width=\textwidth]{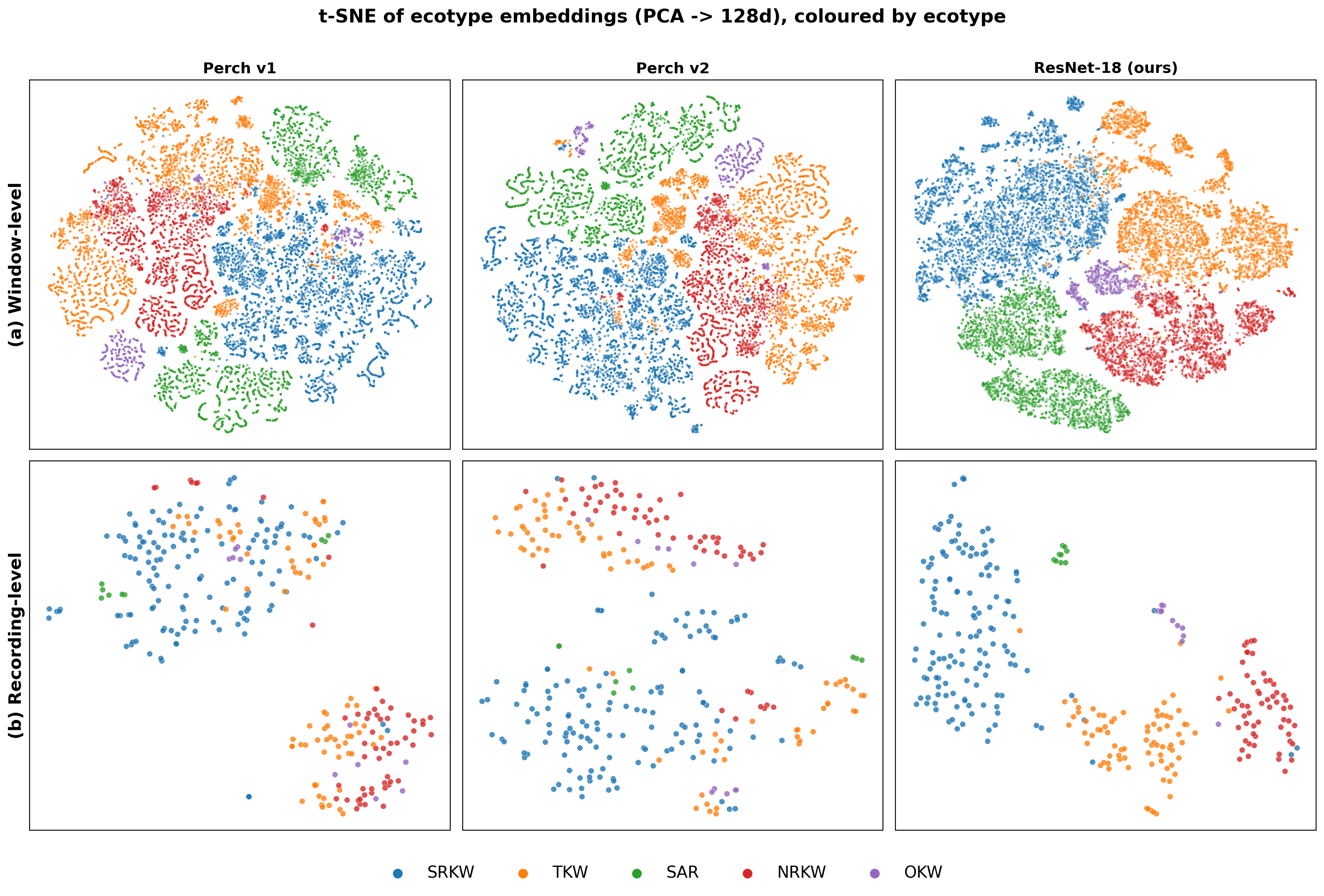}
\caption{t-SNE visualization of ecotype feature embeddings. Embeddings extracted from Perch v1, Perch v2, and the proposed ResNet-18 are projected to two dimensions using t-SNE after PCA reduction to 128 dimensions. Rows show window- and recording-level representations. Points are colored by KW ecotype.}
\label{fig:embeddings}
\end{figure}

Our ecotype classifier outperforms both Perch models at the window level, achieving a macro-F1 of 0.958 compared with 0.933 for Perch 2.0 and 0.913 for Perch 1.0 (\cref{tab:perch}). At the recording level, ResNet-18 reaches a macro-F1 of 0.925 versus 0.873 for Perch 2.0, a gain of 5.2 percentage points. \Cref{fig:embeddings} visualizes the learned representations using t-SNE after PCA dimensionality reduction. Compared with both Perch embeddings, the proposed ResNet-18 produces substantially tighter and better-separated ecotype clusters, particularly after sound file-level aggregation. The qualitative separation closely matches the quantitative improvements reported in \cref{tab:perch}, suggesting that task-specific supervision learns representations that are more coherent for KW ecotype discrimination than general-purpose bioacoustic embeddings.

\subsection{End-to-End Cascade}

\begin{table}[t]
\centering
\caption{Comparison between a single-stage seven-class ResNet-18 and the proposed two-stage cascade on the common held-out test set.}
\label{tab:cascade}
\begin{tabular}{lrr}
\toprule
Class & Single stage F1 & Cascade F1 \\
\midrule
NonBio & 0.960 & 0.965 \\
Bio    & 0.955 & 0.961 \\
SRKW   & 0.930 & 0.932 \\
TKW    & 0.895 & 0.898 \\
SAR    & 0.956 & 0.943 \\
NRKW   & 0.898 & 0.902 \\
OKW    & 0.842 & 0.930 \\
\midrule
Macro-F1 & 0.919 & 0.933 \\
\bottomrule
\end{tabular}
\end{table}

As mentioned before, when evaluated with ground-truth detections, the Stage~2 classifier achieves a macro-F1 of 0.958. Using Stage~1 predictions instead reduces the ecotype macro-F1 to 0.921. Among the 32,033 ecotype-labeled KW windows, 1,590 (5.0\%) are incorrectly classified as Bio or NonBio and therefore do not reach the ecotype classifier, accounting for a 0.020 reduction in macro-F1. False positives introduce a second source of error: 1,273 non-KW windows are incorrectly forwarded to Stage~2 and necessarily assigned one of the five ecotypes, reducing macro-F1 by a further 0.017.

Although abstention is not evaluated as part of the benchmark cascade, applying the calibrated confidence threshold rejects 737 (58\%) of these incorrectly forwarded non-KW windows, preventing them from receiving an erroneous ecotype prediction. This illustrates the complementary roles of the two stages: Stage~1 determines which windows are forwarded for ecotype classification, while confidence-based abstention can limit the propagation of uncertain predictions during deployment.

Despite the performance loss introduced by cascading the two stages, the full seven-class cascade achieves a macro-F1 of 0.933, compared with 0.919 for the single-stage ResNet-18 (\cref{tab:cascade}). The improvement is concentrated almost entirely on the rare OKW class, whose F1 increases from 0.842 to 0.930, while the remaining classes differ by less than 0.013. This behavior reflects the specialization of the second stage on KW vocalizations, allowing the cascade to improve recognition of the rarest ecotype without substantially compromising performance on the more common classes.

\subsection{Stage~1 Deployment-Domain Adaptation}

\Cref{tab:sparrow} summarizes the performance of successive deployed detector versions on the Puget Sound evaluation set. The baseline detector trained exclusively on DCLDE (v1) transfers poorly to the Puget Sound environment, despite its strong performance on the benchmark dataset. Although it detects 75.4\% of KW calls, its precision is only 27.7\%, indicating that nearly three out of every four reported detections are false positives. Such behavior is not functional for a real-time monitoring system, where every alert must be manually reviewed by experts.

The first active learning iteration (v2), incorporating expert-validated Puget Sound detections, increases KW precision to 100\% on the evaluation set but reduces KW recall to 38.5\%. The detector therefore becomes more conservative, substantially reducing false detections at the expense of retaining fewer verified KW candidate windows. The resulting F1 increases from 0.405 to 0.556, reflecting a more balanced precision--recall trade-off for the deployment setting.

The last model (v3) further improves the balance between precision and recall. After incorporating additional manually verified deployment data, ship-noise augmentation, and a reduced fine-tuning learning rate, KW recall increases from 38.5\% to 61.5\% relative to v2, while precision remains at 97.6\%, yielding an KW F1 score of 0.755. 

F1 improves consistently across the three deployment iterations, indicating that iterative fine-tuning can improve detector performance on manually verified candidate windows under the observed domain shift. Further improving recall remains important for conservation of the endangered SRKW population, but identifying hard false negatives requires exhaustive manual verification of the continuous audio stream. Consequently, continuous-stream detection performance and recovery of missed KW vocalizations remain to be evaluated as the deployment progresses.

\begin{table}[t]
\centering
\caption{Detection performance of the base model (v1) and successive active learning iterations (v2, v3) on the Puget Sound data evaluation set.}
\label{tab:sparrow}
\begin{tabular}{lrrrr}
\toprule
Model & Macro-F1 & KW-P & KW-R & KW-F1 \\
\midrule
v1 & 0.445 & 0.277 & 0.754 & 0.405 \\
v2 & 0.511 & 1.000 & 0.385 & 0.556 \\
v3 & 0.579 & 0.976 & 0.615 & 0.755 \\
\bottomrule
\end{tabular}
\end{table}

\section{Conclusion}

We presented a lightweight two-stage cascade for real-time passive acoustic monitoring of KWs, combining a ResNet-18 KW detector with a specialized ecotype classifier. On the DCLDE benchmark, Stage~1 achieves a macro-F1 of 0.960, while Stage~2 reaches 0.958 across five ecotypes, exceeding 0.93 for every class. The ecotype classifier outperforms Perch~2.0 by 2.5 percentage points at the window level and 5.2 points at the recording level, demonstrating the value of task-specific supervision for ecotype recognition. The cascade improves over a single-stage seven-class model (macro-F1 0.933 vs.\ 0.919), with the largest gain for the rare OKW ecotype (0.930 vs.\ 0.842). Error analysis shows that cascade performance is primarily limited by Stage~1 detection errors, while abstention can suppress a substantial fraction of false detections forwarded to the ecotype classifier.

We further adapted the Stage~1 detector to real KW monitoring in Puget Sound under severe domain shift. An active-learning workflow using expert-validated detections and site-specific background noise increased KW precision from 0.28 to 0.98 while improving F1 across successive deployment iterations. Because the ongoing deployment is focused on detecting and alerting on SRKW presence and no SRKW detections have been identified to date, domain adaptation was limited to Stage~1. Future work will include continuous-stream validation to better assess missed detections, extend adaptation and evaluation to Stage~2 and the complete cascade as target-domain ecotype-labeled detections become available, and perform site-specific calibration of the ecotype classification abstention threshold. 

Each stage requires only 11.2M parameters and 5.009 GMACs per 3-second window, with model-only batch-1 inference of approximately 1.4~ms on an NVIDIA H100 NVL, demonstrating faster than real time execution. Although edge-device performance remains to be evaluated, the low model execution cost supports the suitability of the proposed pipeline for real-time passive acoustic monitoring of killer whales in conservation applications.

\bibliographystyle{splncs04}
\bibliography{main}

@misc{noaa2026,
  author = {{NOAA Fisheries}},
  title = {Southern Resident Killer Whale (\textit{Orcinus orca})},
  year = {2026},
  note = {Accessed July 7, 2026},
  url = {https://www.fisheries.noaa.gov/west-coast/endangered-species-conservation/southern-resident-killer-whale-orcinus-orca}
}

@article{scott2024,
  author = {Scott, J. L. and others},
  title = {The WhaleReport Alert System: Mitigating threats to whales with citizen science},
  journal = {Biological Conservation},
  volume = {289},
  pages = {110422},
  year = {2024},
  doi = {10.1016/j.biocon.2023.110422}
}

@inproceedings{he2016,
  author = {He, Kaiming and Zhang, Xiangyu and Ren, Shaoqing and Sun, Jian},
  title = {Deep Residual Learning for Image Recognition},
  booktitle = {2016 IEEE Conference on Computer Vision and Pattern Recognition},
  pages = {770--778},
  year = {2016},
  doi = {10.1109/CVPR.2016.90}
}

@article{bergler2019,
  author = {Bergler, Christian and others},
  title = {{ORCA-SPOT}: An Automatic Killer Whale Sound Detection Toolkit Using Deep Learning},
  journal = {Scientific Reports},
  volume = {9},
  number = {1},
  pages = {10997},
  year = {2019},
  doi = {10.1038/s41598-019-47335-w}
}

@article{vanmerrienboer2026,
  author = {van Merri{\"e}nboer, Bart and Dumoulin, Vincent and Hamer, Jesse and Harrell, Lauren and Burns, Alex and Denton, Tom},
  title = {Perch 2.0: The Bittern Lesson for Bioacoustics},
  journal = {arXiv preprint arXiv:2508.04665},
  year = {2026},
  doi = {10.48550/arXiv.2508.04665}
}

@article{kahl2021,
  author = {Kahl, Stefan and Wood, Connor M. and Eibl, Maximilian and Klinck, Holger},
  title = {{BirdNET}: A deep learning solution for avian diversity monitoring},
  journal = {Ecological Informatics},
  volume = {61},
  pages = {101236},
  year = {2021},
  doi = {10.1016/j.ecoinf.2021.101236}
}

@article{ghani2023,
  author = {Ghani, B. and Denton, T. and Kahl, S. and Klinck, H.},
  title = {Global birdsong embeddings enable superior transfer learning for bioacoustic classification},
  journal = {Scientific Reports},
  volume = {13},
  number = {1},
  pages = {22876},
  year = {2023},
  doi = {10.1038/s41598-023-49989-z}
}

@article{chen2022beats,
  author = {Chen, S. and others},
  title = {{BEATs}: Audio Pre-Training with Acoustic Tokenizers},
  journal = {arXiv preprint arXiv:2212.09058},
  year = {2022},
  doi = {10.48550/arXiv.2212.09058}
}

@article{huang2022,
  author = {Huang, Po-Yao and others},
  title = {Masked Autoencoders that Listen},
  journal = {arXiv preprint},
  year = {2022}
}

@article{chen2024eat,
  author = {Chen, Wenxi and Liang, Yuzhe and Ma, Ziyang and Zheng, Zhisheng and Chen, Xie},
  title = {{EAT}: Self-Supervised Pre-Training with Efficient Audio Transformer},
  journal = {arXiv preprint arXiv:2401.03497},
  year = {2024},
  doi = {10.48550/arXiv.2401.03497}
}

@article{hagiwara2022aves,
  author = {Hagiwara, Masato},
  title = {{AVES}: Animal Vocalization Encoder based on Self-Supervision},
  journal = {arXiv preprint arXiv:2210.14493},
  year = {2022},
  doi = {10.48550/arXiv.2210.14493}
}

@article{rauch2025mae,
  author = {Rauch, Lukas and Heinrich, Robin and Moummad, Ilyass and Joly, Alexis and Sick, Bernhard and Scholz, Christin},
  title = {Can Masked Autoencoders Also Listen to Birds?},
  journal = {arXiv preprint arXiv:2504.12880},
  year = {2025},
  doi = {10.48550/arXiv.2504.12880}
}

@article{zimmermann2026,
  author = {Sch{\"a}fer-Zimmermann, J. C. and others},
  title = {animal2vec and {MEERKAT}: A self-supervised transformer for rare-event raw audio input and a large-scale reference dataset for bioacoustics},
  journal = {Methods in Ecology and Evolution},
  volume = {17},
  number = {3},
  pages = {875--888},
  year = {2026},
  doi = {10.1111/2041-210x.70218}
}

@article{miron2026,
  author = {Miron, M. and others},
  title = {{AVEX}: What Matters for Animal Vocalization Encoding},
  journal = {arXiv preprint arXiv:2508.11845},
  year = {2026},
  doi = {10.48550/arXiv.2508.11845}
}

@article{rauch2025birdset,
  author = {Rauch, Lukas and others},
  title = {{BirdSet}: A Large-Scale Dataset for Audio Classification in Avian Bioacoustics},
  journal = {arXiv preprint},
  year = {2025}
}

@article{hagiwara2022beans,
  author = {Hagiwara, Masato and Hoffman, Benjamin and Liu, Jen-Yu and Cusimano, Mario and Effenberger, Felix and Zacarian, Katie},
  title = {{BEANS}: The Benchmark of Animal Sounds},
  journal = {arXiv preprint arXiv:2210.12300},
  year = {2022},
  doi = {10.48550/arXiv.2210.12300}
}

@article{palmer2025,
  author = {Palmer, K. J. and others},
  title = {A Public Dataset of Annotated \textit{Orcinus orca} Acoustic Signals for Detection and Ecotype Classification},
  journal = {Scientific Data},
  volume = {12},
  number = {1},
  pages = {1137},
  year = {2025},
  doi = {10.1038/s41597-025-05281-5}
}

@article{palmer2026,
  author = {Palmer, K. J. and others},
  title = {Population-Level Acoustic Classification of Salish Sea Killer Whales: Integrating Biologically Informed Call-Type Balancing to Build Robust Models for Conservation Monitoring},
  journal = {Marine Mammal Science},
  volume = {42},
  number = {1},
  pages = {e70126},
  year = {2026},
  doi = {10.1111/mms.70126}
}

@article{burns2025,
  author = {Burns, Alex and Harrell, Lauren and van Merri{\"e}nboer, Bart and Dumoulin, Vincent and Hamer, Jesse and Denton, Tom},
  title = {Perch 2.0 transfers ``whale'' to underwater tasks},
  journal = {arXiv preprint arXiv:2512.03219},
  year = {2025},
  doi = {10.48550/arXiv.2512.03219}
}

@article{castellote2026,
  author = {Castellote, Manuel and others},
  title = {Adaptive Acoustic Monitoring for Endangered Cook Inlet Beluga Whales in Complex Soundscapes},
  journal = {Marine Mammal Science},
  volume = {42},
  number = {3},
  pages = {e70213},
  year = {2026},
  doi = {10.1111/mms.70213}
}

@article{williams2025using,
  title={Using tropical reef, bird and unrelated sounds for superior transfer learning in marine bioacoustics},
  author={Williams, Ben and Van Merri{\"e}nboer, Bart and Dumoulin, Vincent and Hamer, Jenny and Fleishman, Abram B and McKown, Matthew and Munger, Jill and Rice, Aaron N and Lillis, Ashlee and White, Clemency and others},
  journal={Philosophical Transactions of the Royal Society B: Biological Sciences},
  volume={380},
  number={1928},
  pages={20240280},
  year={2025}
}

@article{chasmai2026metaperch,
  title={MetaPerch: Learning from metadata for bioacoustics foundation models},
  author={Chasmai, Mustafa and Dumoulin, Vincent and Hamer, Jenny},
  journal={arXiv preprint arXiv:2607.14072},
  year={2026}
}

\end{document}